\documentclass[fleqn,usenatbib]{mnras}

\usepackage[T1]{fontenc}
\usepackage{ae,aecompl}
\usepackage{soul}

\usepackage{graphicx}	
\usepackage{amsmath}	
\usepackage{amssymb}	
\usepackage[dvipsnames]{xcolor}

\newcommand{\Euclid}{{\em Euclid}}

\newcommand{\Rubin}{{\em Rubin}}

\usepackage{newtxtext,newtxmath}

\title[Individual weak lensing masses]{Measuring weak lensing masses on individual clusters}

\author[C. Murray et al.]{
Calum Murray,$^{1,2}$\thanks{E-mail: calumhrmurray@gmail.com}
James G. Bartlett,$^{1,3}$
Emmanuel Artis,$^{4,2}$
\newauthor
Jean-Baptiste Melin$^{4}$
\\
$^{1}$Université de Paris, CNRS, Astroparticule et Cosmologie, F-75013 Paris, France\\
$^{2}$Université Grenoble Alpes, CNRS, LPSC-IN2P3, 38000 Grenoble, France\\
$^{3}$Jet Propulsion Laboratory, California Institute of Technology, 4800 Oak Grove Drive, Pasadena, CA, 91109, USA\\
$^{4}$IRFU, CEA, Universit{\'e} Paris-Saclay, F-91191 Gif-sur-Yvette, France
}

\date{Accepted XXX. Received YYY; in original form ZZZ}

\pubyear{2021}

\begin{document}
\label{firstpage}
\pagerange{\pageref{firstpage}--\pageref{lastpage}}
\maketitle

\begin{abstract}
We present weak lensing mass estimates for a sample of 458 galaxy clusters from the redMaPPer Sloan Digital Sky Survey DR8 catalogue using Hyper Suprime-Cam weak lensing data. We develop a method to quickly estimate cluster masses from weak lensing shear and use this method to estimate weak lensing masses for each of the galaxy clusters in our sample. Subsequently, we constrain the mass-richness relation as well as the intrinsic scatter between the cluster richness and the measured weak lensing masses. When calculating the mass-richness relation for all clusters with a richness $\lambda>20$, we find a tension in the slope of the mass-richness relation with the Dark Energy Survey Year 1 stacked weak lensing analysis. For a reduced sample of clusters with a richness  $\lambda>40$, our results are consistent with the Dark Energy Survey Year 1 analysis.
\end{abstract}




\section{Introduction}

Galaxy clusters are amongst the largest and most recently formed structures in the universe. As initially small density fluctuations grow through gravitational instability, the largest regions collapse and decouple from the background expansion of the universe forming galaxy clusters. Measurements of the masses and abundance of galaxy clusters provide information on the expansion history of the universe, its geometry and also the nature of gravity \citep{allen2011cosmological}. 

Much of the cosmological information in cluster abundance cosmology relies on a robust estimation of cluster masses. Weak gravitational lensing is an important tool to estimate cluster masses as it allows us to directly access the projected surface mass density of clusters largely independently of their dynamical state or through assumptions on complex baryonic physics \citep{umetsu2020cluster}.

In this work we estimate cluster masses for a subsample of the redMaPPer Sloan Digital Sky Survey (SDSS) cluster catalog \citep{rykoff2014redmapper} using weak lensing data from the Hyper Suprime-Cam survey \citep{aihara2018hyper}. Masses are estimated for individual clusters using the excess surface mass density profiles and a matched filter method. Subsequently, we obtain the mass-richness relation for our measurements and find a shallower slope than the stacked analysis by the Dark Energy Survey (DES) \citep{mcclintock2019dark}. We then show that there are important differences between stacked mass estimates and individual mass estimates. These differences are not, however, sufficient to resolve the disagreement between the analyses. Such effects will be important for future high precision cluster surveys, such as \Rubin\  \citep{ivezic2019lsst}  and \Euclid\ \citep{laureijs2011euclid}.

In Sect.~2 we review the basics of weak gravitational lensing for galaxy clusters and present our method for estimating cluster masses from the excess surface density profile around galaxy clusters. In Sect.~3 we present the data used in our analysis and the joint distribution of cluster richness and weak lensing mass. In Sect.~4 we constrain the mass-richness relation and compare it to other results in the literature. Subsequently, in Sect.~5 we compare our mass estimates to that of \cite{umetsu2020weak}. We discuss the differences between stacked analysis and individual mass estimates in Sect.~6 before concluding in Sect.~7.

\section{ Method }

\subsection{ Weak lensing around clusters }

Weak gravitational lensing around galaxy clusters distorts the shapes of background galaxy images such that the observed ellipticity is the sum of the intrinsic ellipticity of the galaxy and the shear induced by weak lensing.
\begin{equation*}
    (\epsilon_{\rm obs,1}, \epsilon_{ \rm obs,2}) = (\epsilon_{ \rm int,1}, \epsilon_{ \rm int,2}) + ( \gamma_1 , \gamma_2 ).
\end{equation*}
Under the assumption that there is no spatial correlation between the intrinsic shapes of the galaxies, a reasonable assumption when the physical separation between galaxies is large, the average of the galaxy shapes over some sky area provides a measure of the shear, $\gamma$.

For a spherically symmetric lens, we have only a tangential component of shear, $\gamma_{+}$. The radial dependence of the tangential shear can be written as,
\begin{equation*}
\gamma_{+}(R) = \frac{\Delta \Sigma(R)}{\Sigma_{\rm crit}(z_l,z_g)},
\end{equation*}
where we have introduced the excess surface mass density, $\Delta \Sigma(R) =\bar{\Sigma} (R) - \Sigma (R)$, with $\Sigma$ the surface mass density (the mass density integrated along the line-of-sight) and  $\bar{\Sigma} (R)$ the average surface mass density interior to $R$ \citep{miralda1991gravitational}.

The magnitude of the weak lensing shear depends on both the source and lens redshifts. As such it is useful instead to work with the excess surface density, $\Delta \Sigma(R)$, which is independent of source redshift, by appropriately weighting the measured galaxy ellipticities with $\Sigma_{\rm crit}(z_l,z_g)$, the critical surface mass density,

\begin{equation*}
\Sigma_{\rm crit}(z_l,z_g) \equiv \frac{c^2}{4 \pi G} \frac{D_s}{D_l D_{ls}}
\end{equation*}

where $D_s$, $D_l$ and $D_{ls}$ are the angular diameter distance to the source, lens and between the lens and source respectively.  Here, $z_l$ is the lens redshift and $z_g$ is the redshift of a background galaxy. 

\subsection{Mass measurement method}
\label{mass_measurements:section:mass_filter}

The measured excess surface density at  radius $r_i$ from a cluster center can be modelled as the signal, $\Delta \Sigma(r_i)$, contaminated by some noise, $n(r_i)$, due primarily to the intrinsic source ellipticity.
\begin{equation}
    \Delta \Sigma_i = \Sigma_{crit}(z_l,z_g)( \gamma_{+}(r_i) + \epsilon_{int,+}(r_i)) = \Delta \Sigma(r_i) + n(r_i)
\end{equation}{}
It is instructive to rewrite the above equation in the following form. We want to calculate the amplitude of the lensing signal $a$ given a known template for the signal $\tau(r_i)$,
\begin{equation}
\label{eq:mass_measurements:signal}
    \Delta \Sigma_i = a \tau(r_i) + n(r_i)
\end{equation}{}

The template we use is the NFW density profile such that, $\Delta \Sigma(r_i) = r_s \tau( r_i ) = a \tau( r_i )$. Where $r_s$ is the scale radius of the NFW profile from which we can obtain the lens mass and $\tau$ is (\cite{wright2000gravitational}),
\begin{equation}
    \tau(x)  = 
    \begin{cases}
      \frac{ \delta_c \rho_c}{ ( x^2-1 )} 
       [ \frac{8}{x^2\sqrt{1-x^2}} \text{ arctanh } \sqrt{ \frac{1-x}{1+x} } 
        + \frac{4}{x^2} \text{ ln } \frac{x}{2}  
        - \frac{2}{x^2 - 1 } \\
        + \frac{4}{(x^2-1)\sqrt{1-x^2}} \text{ arctanh } \sqrt{ \frac{1-x}{1+x} } ] 
      \text{ for } x < 1 \\
      
       \frac{ \delta_c \rho_c}{3 } [ \frac{10}{3} + 4 \text{ ln } \frac{1}{2}]
       \text{ for } x = 1 \\
       
       \frac{ \delta_c \rho_c}{( x^2-1 )} 
       [ \frac{8}{x^2\sqrt{1-x^2}} \text{ arctan } \sqrt{ \frac{x-1}{1+x} } 
        + \frac{4}{x^2} \text{ ln } \frac{x}{2} 
        - \frac{2}{x^2 - 1 } \\
        + \frac{4}{(x^2-1)\sqrt{1-x^2}} \text{ arctan } \sqrt{ \frac{x-1}{1+x} }  ]  
       \text{ for } x > 1 
    \end{cases}   
\label{5:eq:shear_nfw}
\end{equation}  
with $x=r/r_s$. With this data model, we construct a linear filter, $h$, to maximise the output signal-to-noise ratio. The filter output $a_{est}$ is the inner product of the filter and the observed signal $\Delta \Sigma_i$:
\begin{equation}
    a_{est} = \sum_i \Delta \Sigma_i h_i
\end{equation}
where $a_{est}$ corresponds to the amplitude of the signal, and by construction $a_{est} = r_{s,est}$; therefore, we can infer the lens mass from the amplitude of the filter output with $M_l = \frac{4}{3} \Delta \rho_c ( r_s c )^3$, where we must assume a concentration, $c$, and choose an overdensity, $\Delta$. 

The two conditions that we wish our filter to satisfy are that it is unbiased, $b=0$, and that the noise, $\sigma$, is minimal,

\begin{equation}
    b = <a_{est}-a> = \sum_i \langle\Delta \Sigma_i h_i \rangle - a = a \left ( \sum_i \tau_i h_i  - 1 \right ) + \left < \sum_i n_i h_i \right >
\end{equation}

\begin{equation}
    \sigma^2 = <(a_{est}-a)^2> = \left( \boldsymbol{\tau}^T \boldsymbol{S}^{-1} \boldsymbol{\tau} \right)^{-1}
\end{equation}

where $\boldsymbol{S}$ is the covariance matrix of the noise and $\boldsymbol{\tau}$, $\boldsymbol{\tau}^T$ are row and column vectors respectively. To find the filter, $h$, which satisfies these two conditions, we vary the Lagrangian $L = \sigma^2 + \lambda b$ with respect to $h_i$, which gives the filter,

\begin{equation}
    \boldsymbol{h} = \frac{ \boldsymbol{\tau}^T \boldsymbol{S}^{-1} }{   \boldsymbol{\tau}^T \boldsymbol{S}^{-1} \boldsymbol{\tau} }
\end{equation}
With our matched filter, $h_i$, we can calculate this value $r_s$ given a template for the cluster lensing signal and the noise covariance matrix,
\begin{equation}
    r_s = a = \frac{ \sum \Delta \Sigma_i h_i }{ \sum \tau(r_i) h_i }.
\end{equation}{}

 As the template of the matched filter is informed by both the lens mass and concentration, we obtain our mass estimates by minimising the square of the difference between filter output and input,
 \begin{equation}
    f = \left( \sum_i h_i \Delta \Sigma_i - r_s \sum_i h_i \tau_i \right)^2
\end{equation}
We use the mass-concentration as measured in \citep{bhattacharya13} and use an overdensity $\Delta = 200 \rho_m$.

\section{ Data and measurements }

\subsection{ SDSS RedMaPPer Catalog }

We used the public redMaPPer catalogue  \footnote{ http://risa.stanford.edu/redmapper/ } based on SDSS-DR8, which contains 26,111 clusters \citep{rykoff2014redmapper}. The redMaPPer cluster finding algorithm was applied to the Sloan Digital Sky Survey (SDSS), a large-scale galaxy survey that mapped over a million galaxies across half of the northern sky \citep{aihara2011erratum,york2000sloan}. SDSS has been conducted with a 2.5m wide-angle optical telescope at the Apache Point Observatory in New Mexico. The redMaPPer cluster finder was run on 10,400~${\rm deg}^2$ of photometric data from the 8th data release. Each cluster has an assigned brightest cluster galaxy, which is used as the cluster centre, as well as right ascension, declination, redshift and a list of the cluster member galaxies with a membership probability for each. The catalog consists of clusters in the redshift range $ z \in [ 0.08 , 0.6 ]$ and the richness range $\lambda \in [ 20 , 300 ]$, where the cluster richness is defined as the sum over the membership probabilities of galaxies for a given cluster. 

\subsection{ Hyper Suprime-Cam shear catalog }

In addition to the redMaPPer clusters, we use galaxy shape measurements and photometric redshifts from the first year data of the Hyper Suprime-Cam (HSC) Subaru Strategic Program to calculate the lensing masses of the clusters \citep{aihara2018hyper}. Hyper Suprime-Cam is a digital imaging camera operating on the 8.2m Subaru telescope in Hilo, Hawaii.

\begin{figure}
\centering
\includegraphics[width=9cm]{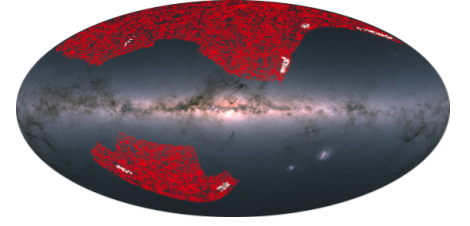}
\caption{ Red points show SDSS RedMaPPer clusters. White points show these clusters for which we also have weak lensing mass estimates using HSC data. The image is projected onto the Gaia DR2 map \citep{brown2018gaia}. }
\label{image:cluster_catalog_overview}
\end{figure}

The HSC-Wide shear catalog used in this analysis covers 137~${\rm deg}^2$ of the sky. The overlap with the redMaPPer SDSS-DR8 catalog is shown in Figure \ref{image:cluster_catalog_overview}. Clusters from the redMaPPer catalogue were selected within the HSC survey area, where their fields of view have no significant masked regions. This leaves 458 galaxy clusters for which we have sufficient lensing information to calculate the cluster lensing masses. 

\subsection{ Source galaxy selection }

We use the photometric redshifts obtained with the point-spread function matched aperture photometry code, Ephor\_AB \citep{tanaka2018photometric}. This is a publicly available photometric redshift estimation code that uses a feed-forward neural network \footnote{ https://hsc-release.mtk.nao.ac.jp/doc/index.php/photometric-redshifts/ }.

To avoid dilution of the lensing signal by cluster member galaxies, we employ a conservative cut on galaxy photo-zs  following \citep[][]{umetsu2020weak},
\begin{equation}
    p_{cut} < \int_{z_{min}}^{\infty} P(z) dz \text{ and } z_p < z_{max}
\end{equation}
where we use $p_{cut} =0.98$ and $z_{min} = z_l + 0.2 $, where $z_l$ is the redshift of the cluster, $z_p$ is a randomly sampled point estimate for the galaxy redshift drawn from its probability distribution $P(z)$ and $z_{max}=2.5$. 

\subsection{ Excess surface density estimation }
\label{mass_measurement:psf}

The principle systematic contaminant to galaxy shape measurements is the effect of the camera point-spread function (PSF) \citep[e.g.,][]{kaiser1994method}. This is the response of the camera to a point source which causes an imaged point source to appear as a blurry image rather than as a sharp point. The PSF creates a multiplicative bias, $m$, to the true shear such that the estimated shear, $\hat{g}$, is related to the true shear, $g$, as
\begin{equation}
    \hat{g} = ( 1 + m ) g
\end{equation}
for the average shear over many galaxies. Shape measurements in the HSC catalog \citep{mandelbaum2018first,bosch2018hyper} are calculated from moments of the galaxy image using the re-Gaussianization technique \citep{hirata2003shear} that corrects for the PSF. The multiplicative bias is then calibrated on image simulations \citep{mandelbaum2018weak}.

The residual multiplicative shear bias for an ensemble of galaxies is calculated as,
\begin{equation}
    1 + K(R_i) = \frac{ \sum_i w_i ( 1 + m_i ) }{ \sum_i w_i}
\end{equation}
where $m_i$ the multiplicative bias for a given galaxy and $w_i$ is the inverse variance weight for a given galaxy, which is a combination of the measurement uncertainty on the shape, $\sigma_{\epsilon,mes}^2$, and the intrinsic shape noise, $\sigma_{\epsilon,int}^2$,
\begin{equation}
    w_i = \frac{1}{\sigma_{\epsilon,int}^2 + \sigma_{\epsilon,mes}^2} \frac{1}{ {\Sigma^{MC}_{crit,i}}^2 }
\end{equation}

where the galaxies are weighted by their lensing strength with the critical surface mass density. Instead of the expectation value of $\Sigma_{crit}$ we use a Monte Carlo estimate where $z^{MC}_{s}$ is a random sample from $p_{phot}(z_s)$ which is provided by the HSC data, following the approach of \cite{mcclintock2019dark}.

In addition, ensemble shear measurements should be corrected for the shear responsivity. This is the response of an intrinsic ellipticity to a small shear. It is calculated as,
\begin{equation}
    \mathcal{R}(R_i) = 1 - \frac{ \sum_i w_i \epsilon^2_{int} }{ \sum_i w_i }
\end{equation}
Where in general $\epsilon_{int} \approx 0.4$ and $\mathcal{R} \approx 0.84$ \citep{mandelbaum2005systematic,mandelbaum2018first}. 

The excess surface mass density in radial bins for individual clusters is computed as \citep{mandelbaum2018first},
\begin{equation}
\label{eq:mass_measurements:esd_estimator}
\Delta \Sigma_j = \frac{1}{2 \mathcal{R}_j } \frac{1}{1+K_j} \left[ \frac { \sum_i w_i \epsilon_{+,i} \Sigma^{MC}_{crit,i} } {\sum_i w_i} \right ]_j
\end{equation}

where the index $i$ refers to the sum over individual galaxies in a radial bin $j$.

In addition, the effective bin radius is calculated from the weighted harmonic mean of the lens-source transverse separations following \cite{okabe2016locuss} and \cite{umetsu2020weak} as,
\begin{equation}
    R_i = \frac{ \sum_i w_i}{ \sum_i w_i R^{-1}_i }.
\end{equation}

These distances are subsequently used as the input into the matched filter algorithm. In practice, the effective bin radius calculated with this method does not vary significantly from using the centre of the radial bins.

We require a covariance matrix for each cluster in order to make our mass measurements.  Our covariance matrices are calculated from the lensing weights as,
\begin{equation}
    \sigma^2(R_i) = \frac{1}{4 \mathcal{R}^2(R_i)} \frac{1}{\left[ 1 + K(R_i) \right]^2 \sum_i w_i }.
\end{equation}

Other noise contributions to the lensing profiles such as from correlated large-scale structure, uncorrelated structures along the line of sight or the intrinsic variance of cluster excess surface mass density profiles at a given cluster mass our included at the level of estimation of the mass-richness relation, as explained in the following section. As such our approach is to explicitly estimate the cluster weak lensing mass and its error, as opposed to the uncertainity between the `true' cluster mass and our  estimate.

In order to maintain the simple relation between the cluster mass and excess surface mass density signal, the filter must be applied over a radial range where the cluster mass dominates the signal. At small radii cluster miscentering and concentration are important, and at large radii the signal from nearby correlated structure is important. In order to minimise such contributions, we perform the mass estimation over the radial range $R \in [0.8,3.54]$ Mpc with logarithmically spaced bins of size $\Delta \log_{10} R = 0.13$. We tested both changing the mass-concentration relation and the number of radial bins and found that the changes were neglible to the resultant estimated weak lensing masses.

\subsection{ Mass estimates }
\begin{figure}
\centering
\includegraphics[width=8cm]{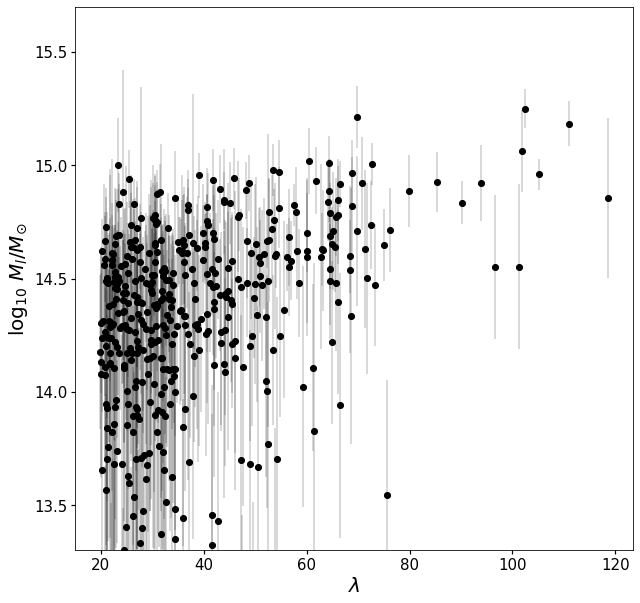}
\caption{ Lensing masses have been estimated using our method and HSC data for 458 indvidual galaxy clusters. The richnesses are taken from the redMaPPer SDSS DR8 catalogue. The grey error bars show the $1 \sigma$ error on each mass measurement. }
\label{image:mass_measurement:scatter}
\end{figure}

\begin{figure}
\centering
\includegraphics[width=9cm]{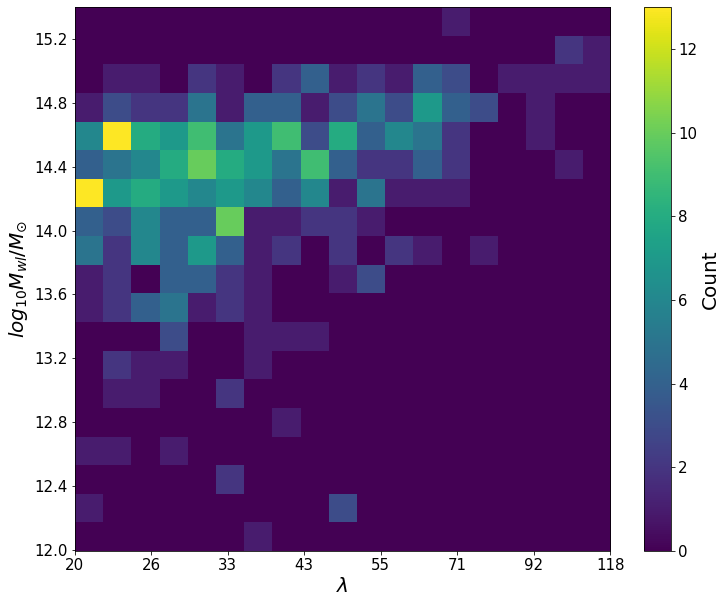}
\caption{ The joint distribution of lensing masses and richness. Lensing masses have been estimated using our method and HSC data. The richnesses are taken from the redMaPPer SDSS DR8 catalogue. Whilst noisy for this small sample, future cluster surveys will allow this joint distribution to constrain the correlations between lensing masses and richness.}
\label{image:mass_measurement:joint_richness_mass_distribution}
\end{figure}

The lensing masses can now be estimated using our method as outlined in Section \ref{mass_measurements:section:mass_filter}. The analysis is performed at a fixed cosmology, assuming $\Lambda$CDM with $\Omega_m = 0.3$, $\Omega_{\Lambda} = 0.7$ and $H_0 = 70$ [km/s/Mpc]. The errors on each mass estimate are calculated by bootstrap resampling of the background galaxies for each cluster. In Figure \ref{image:mass_measurement:scatter} we show the obtained lensing masses and their richness. The joint distribution of lensing masses and richness is shown in Figure \ref{image:mass_measurement:joint_richness_mass_distribution}. 

\section{ Mass-richness relation }
We now have a sample of clusters with weak lensing mass and richness measurements, each with attributed uncertainties. At this point we can look at the relation between cluster mass and  cluster richness.  We use the same power-law relation used in \cite{mcclintock2019dark},
\begin{equation}
\label{eq:mr_relation}
    \left< M_{wl} | \lambda ,z \right > = 10^{M_0} \left( \frac{\lambda}{\lambda_0}\right)^{F_{\lambda}} \left( \frac{1+z}{1+z_0}\right)^{G_{z}}
\end{equation}
where $M_0$, $F_{\lambda}$ and $G_z$ are parameters to be estimated, and we set $\lambda_0=40$ and $z_0=0.35$, which are approximately the median values for each property in our sample and the same values as used in \cite{mcclintock2019dark}. Additionally, as we have weak lensing mass measurements for each cluster, we can constrain the intrinsic scatter, $\sigma_{\ln M_{wl},\rm{int}|\lambda}$, in the measured weak lensing mass at given cluster richness. 

\begin{figure}
\centering
\includegraphics[width=9cm]{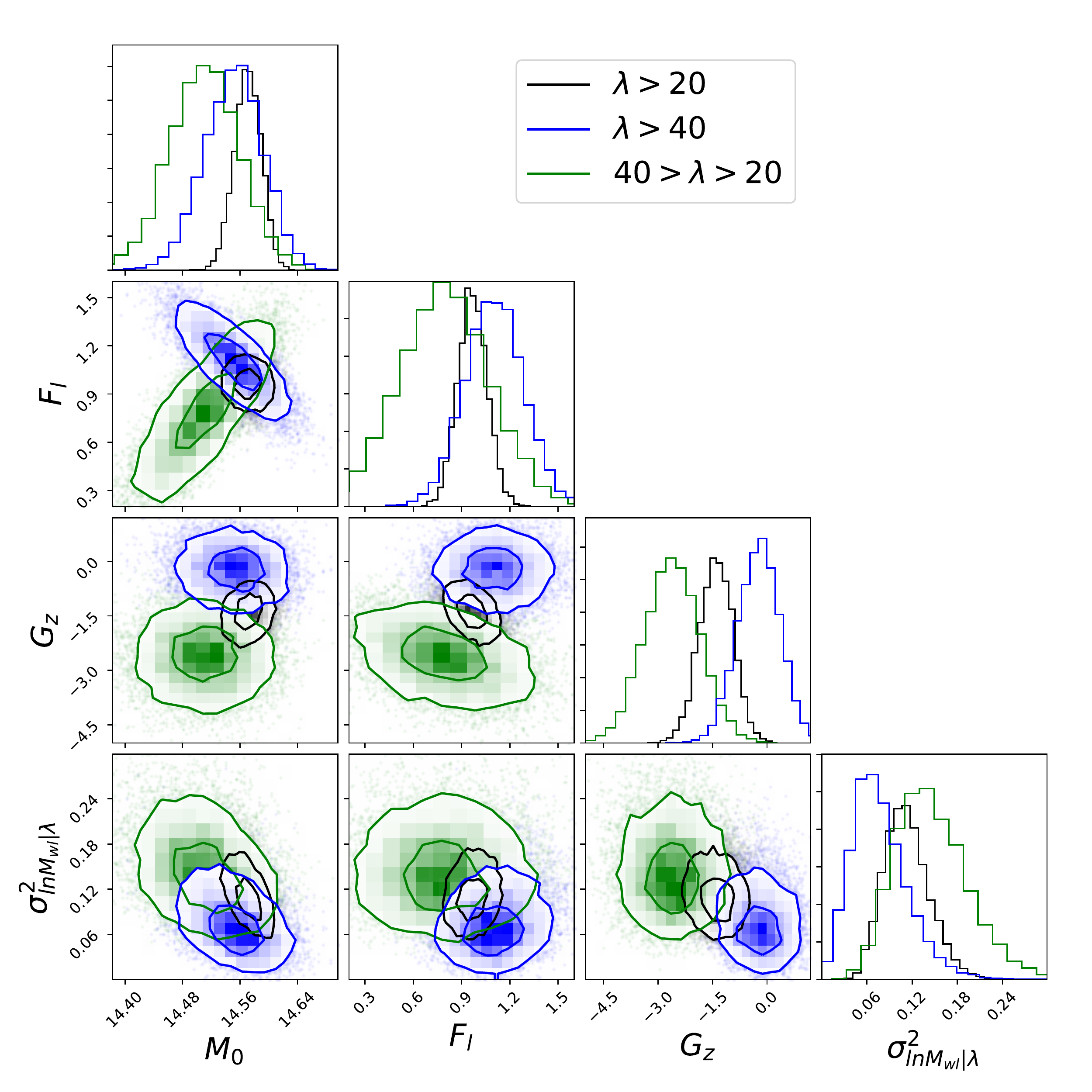}
\caption{ Posteriors for the parameters of the mass-richness relation in Equation \ref{eq:mr_relation}. Contours show the $1 \sigma$ and $2 \sigma$ confidence areas. This figure was produced using the python package corner \protect\cite{corner}.}
\label{image:mass_measurement:mass_richness_mcmc}
\end{figure}

The posterior distribution of the model parameters is calculated using the Markov Chain Monte Carlo (MCMC) technique and the emcee package \citep[][]{foreman2013emcee}. Assuming a normal distribution our likelihood is,
\begin{equation}
    \ln \mathcal{L} = -\frac{1}{2} \sum_i ( \Delta \ln M_i )^2 / \sigma^2_{\ln M_{wl},i} + \ln(\sigma^2_{\ln M_{wl},i})
\end{equation}
where $\Delta \ln M_i = \ln M_i - \ln \left< M | \lambda ,z , M_0 , F_{\lambda} ,G_z \right > $, $\ln M_i$ is the measured weak lensing mass for each cluster and $\sigma^2_{\ln M_{wl},i} =\sigma^2_{\ln M_{wl},\rm{int}|\lambda}+ \sigma^2_{\ln M_{wl},\rm{meas}}$ is the sum in quadrature of the intrinsic, $\sigma_{\ln M_{wl},\rm{int}|\lambda}$, and measurement induced scatter, $\sigma_{\ln M_{wl},\rm{meas}}$ . The summation is over all the clusters in the sample. The results of the MCMC are shown in Figure \ref{image:mass_measurement:mass_richness_mcmc}. 

Using all 458 galaxy clusters with a redMaPPer richness greater than 20, we obtain a mass parameter $M_0 = 14.546 \substack{+0.019 \\ -0.019}$, a richness scaling of $F_l = 0.948 \substack{+0.092 \\ -0.090}$ and a redshift evolution of $G_z=-1.313\substack{+0.471 \\ -0.447}$, and an intrinsic scatter $\sigma^2_{lnM_{wl}|\lambda}=0.118\substack{+0.033 \\ -0.027}$. In calculating the mass parameter, $M_0$, we have corrected for the difference in redMaPPer estimated richness between DES and SDSS following \cite{mcclintock2019dark} using their Eqs. (67) and (69). The richness definition depends on the magnitudes, source detection algorithms and other effects which vary from survey to survey. 

 Additionally, to consider richness dependent effects we create two sub-samples of clusters and repeat the analysis. Firstly a low richness sub-sample with $40 > \lambda > 20$ which contains 299 galaxy clusters. For this sub-sample we obtain a mass parameter $M_0 = 14.478 \substack{+0.051 \\ -0.052}$, a richness scaling of $F_l = 0.726 \substack{+0.290 \\ -0.294}$ a redshift evolution of $G_z=-2.357\substack{+0.759 \\ -0.771}$, and an intrinsic scatter $\sigma^2_{lnM_{wl}|\lambda}=0.177\substack{+0.055 \\ -0.046}$.

 Secondly a high richness sub-sample with $\lambda > 40$ which contains 159 galaxy clusters. For this subsample we obtain a mass parameter $M_0 = 14.528 \substack{+0.037 \\ -0.040}$, a richness scaling of $F_l = 1.095 \substack{+0.178 \\ -0.172}$ a redshift evolution of $G_z=-0.497\substack{+0.602 \\ -0.592}$, and an intrinsic scatter $\sigma^2_{lnM_{wl}|\lambda}=0.061\substack{+0.036 \\ -0.026}$. 
 
 For both the low richness sample, $40 > \lambda > 20$, and the full cluster sample we find a non-zero redshift evolution of the mass richness relation at 3.1$\sigma$ and 2.8$\sigma$ respectively. It is unclear if such an evolution is physical or due to difference in the richness definition at differing redshifts from galaxy member selection effects. 
 
 We see a slight difference at 1.8$\sigma$ between the intrinsic scatter measured on the low richness and high richness sample but our measurements are not yet precise enough to determine if this is a physical effect or a statistical fluctuation. 

 In Figure~\ref{image:mass_measurement:mass_richness_mcmc}  we show the results of the MCMC analysis for each sample of clusters. The results of each analysis, for the three different richness cuts, are in good agreement for both the mass parameter, $M_0$, and the richness scaling of the relation, $F_l$, although with a notable shift in the redshift evolution of the relation, $G_z$, between the low and high richness samples.  

\begin{figure}
\centering
\includegraphics[width=9cm]{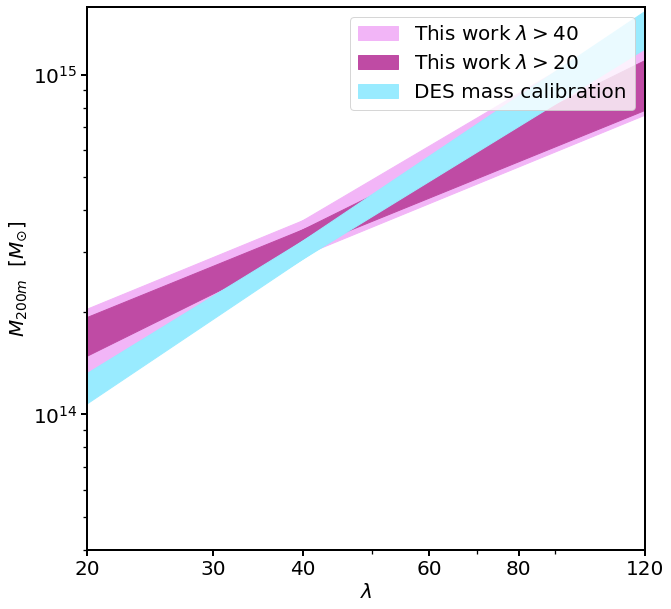}
\caption{ The mass-richness relation from individual lensing masses estimated for all clusters with richness greater than 20 and 40. The DES mass calibration referred to here is \protect\cite{mcclintock2019dark}, where they use clusters with a richness greater than 20.}
\label{image:mass_measurement:mass_richness_relation}
\end{figure}

In Figure \ref{image:mass_measurement:mass_richness_relation} we compare our results to the DES stacked lensing mass results. DES obtain a slightly lower mass parameter $M_0 = 14.489 \pm 0.029$, which is a disagreement of $1 \sigma$ from our own. The DES mass-richness relation is constrained for clusters with a richness, $\lambda>20$. DES find a notably steeper richness scaling of $F_l = 1.356 \pm 0.059$, in  tension with our result at the $3.7 \sigma$ level; this disagreement is substantially reduced to $1.4 \sigma $ when our analysis is restricted to clusters with a richness $\lambda>40$. For the redshift evolution of the relation, DES find $G_z=-0.3 \pm 0.36$, in disagreement with our result at the $2.7 \sigma$ level. As this is a stacked analysis, it is not straightforward to constrain the intrinsic scatter between richness and lensing mass therefore no constraints on the scatter are provided in this analysis.

These differences in the richness scaling could be driven by the different radial ranges in the analyses. Our inner and outer radial cuts are much more conservative than the DES analysis, a necessity given our matched filter template. The differences between the two analyses appear to be driven by low richness clusters.

There are many other constraints on the mass-richness relation of redMaPPer clusters in the literature, a good summary of these results are shown in Figure 15 of \cite{mcclintock2019dark}, our normalisation is in good agreement with the majority of the results. The richness scaling of our mass-richness relation is generally on the lower side although two other analyses in the literature find a shallower slope ( \cite{geach2017cluster}, \cite{saro2015constraints}).

\section{Comparison with Umetsu et al. (2020)}

We compare our results to the weak lensing mass estimates on individual cluster made by \cite{umetsu2020weak}, which is a weak lensing analysis of X-ray galaxy groups and clusters selected from the XMM-XXL survey and uses the same weak lensing HSC data as used in our analysis. In the \cite{umetsu2020weak} analysis, cluster masses are estimated by fitting the tangential excess surface density profile over the radial
range $R \in [0.3, 3]$ $h^{-1}$ Mpc in comoving length units with an NFW profile, leaving both the cluster mass and concentration as free parameters. The \cite{umetsu2020weak} analysis obtains masses defined as $M_{200c}$, therefore we convert our $M_{200m}$ mass measurements to $M_{200c}$ using the code COLOSSUS \cite{colossus} \footnote{https://bitbucket.org/bdiemer/colossus/} and the mass-concentration relation from \cite{bhattacharya13}.

\begin{figure}
\centering
\includegraphics[width=9cm]{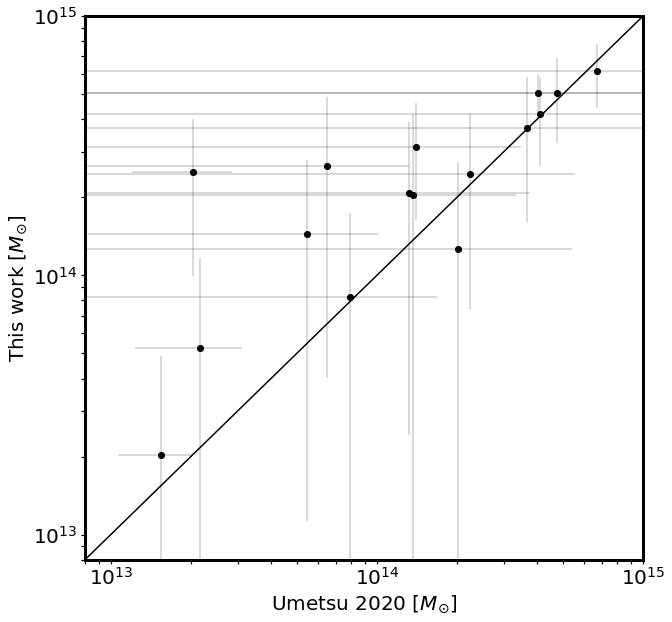}
\caption{ Comparison between  weak lensing mass estimates of \protect\cite{umetsu2020weak} and this work. There are 16 clusters in common between these two samples. We find the two samples to be consistent, with $\left< M_{\rm Umetsu}/ M_{\rm Murray} \right> = 0.78 \pm 0.36 $, although the sample is small and there is large scatter on the indivudal mass estimates.}
\label{image:mass_measurement:hsc_comparison}
\end{figure}

We find 16 clusters in common for this comparison. The scatter on both sets of measurements is considerable as the cluster masses are low. The two samples are consistent with, $\left< M_{\rm Umetsu}/ M_{\rm Murray} \right> = 0.78 \pm 0.36 $. The comparison of these results is shown in Fig. \ref{image:mass_measurement:hsc_comparison}.

\section{ Comparing stacked results to individual mass estimates }

In the DES analysis \citep{mcclintock2019dark}, as is normal in a stacked analysis, the mass inferred from the stacked profile is taken to be equal to the mean mass of the stack weighted by the sum of lens-source pairs for each cluster. This is not strictly true as the relationship between the excess surface mass density and the cluster mass is not linear. Essentially it is not true that,
\begin{equation}
\label{ineq}
\left< \Delta \Sigma | \ln \lambda \right > \neq \Delta \Sigma ( \left< \ln M | \ln \lambda \right>)
\end{equation}
In the following we calculate the impact of such an assumption, which we find to be small, and compare to the case of mass estimates on individual clusters. In order to understand the relation between the mean mass of the stack and the mass inferred by the mean lensing profile of the stack, we model the richness-mass and lensing mass-mass relations as a power-laws with log-normal scatter.
\begin{equation}
    \left< \ln \lambda | M , z \right> = \pi_{\lambda} + \alpha_{\lambda} \ln \left( \frac{M}{M_p}\right )
\end{equation}
\begin{equation}
    \left< \ln M_l | M , z \right> = \pi_{M_l} + \alpha_{M_l} \ln \left( \frac{M}{M_p}\right )
\end{equation}
where $M_p$ is a halo pivot mass scale which may be chosen, $\alpha$ is the slope of the relation and $\pi$ is the normalisation. Here, we have written the power-laws for just two cluster observables; however, these simple relations are justifiable for all cluster mass observables \citep{kaiser1986evolution}. One of the strengths of weak lensing mass estimates is that we can reasonably expect that $\pi_{M_l} \approx \ln M_p$, $\alpha_{M_l} \approx 1$. 

The scatter on these relations can be approximated as log-normal with the joint probability distribution described by the multivariate normal distribution $P \left (\ln M_l, \ln \lambda |  \ln \left( \frac{M}{M_p}\right ) \right )$ with the covariance matrix,
\begin{equation}
    S = 
    \begin{pmatrix}
    \sigma^2_{\ln M_l} & r \sigma_{\ln \lambda} \sigma_{\ln M_l} \\
    r \sigma_{\ln \lambda} \sigma_{\ln M_l}  & \sigma^2_{\ln \lambda}
    \end{pmatrix}
\end{equation}
where $\sigma^2_{\ln M_l}$ and $\sigma^2_{\ln \lambda}$ are the intrinsic log variances of the observables and $r$ is the correlation between them. We can see that the mass-observable relations have seven free parameters, $\pi_{\lambda}$, $\pi_{M_l}$, $\alpha_{\lambda}$, $\alpha_{M_l}$, $ \sigma_{\ln \lambda}$ , $ \sigma_{\ln M_l}$ and $r$.

These relations give the probability distributions we need to calculate each side of Eq. (\ref{ineq}). The stacked excess surface density profile for a stack in richness can be calculated as,
\begin{equation}
\label{mean_profile}
\left<\Delta \Sigma (R_i) | \ln \lambda \right> 
= \int d \ln M   \Delta \Sigma ( R_i | \ln M , c ) P( \ln M | \ln \lambda ) 
\end{equation}

where $P(\ln M | \ln\lambda )$ is calculated from the halo mass function $P(\ln M)$ with,
\begin{equation}
P(\ln M | \ln\lambda ) = P( \ln \lambda | \ln M ) P( \ln M )  / P( \ln \lambda ) 
\end{equation}

The cluster concentration is taken to be independent of the observed cluster richness. 

\begin{figure}
\centering
\includegraphics[width=9cm]{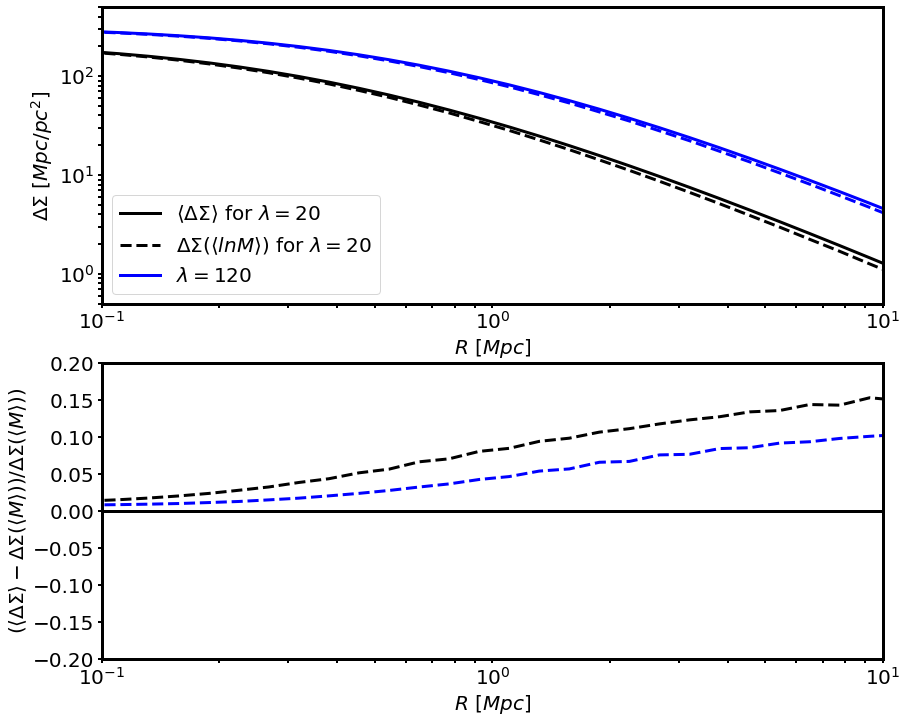}
\caption{ Excess surface density profile for the average mass in the stack $\left< \Delta \Sigma \right>$ and the average excess surface density profile for the average mass of clusters in the stack $ \Delta \Sigma ( \left< \ln M \right > )$. For these calculations we use the \protect\cite{tinker2010large} halo mass function as implemented in hmfcalc \protect\citep{murray2013hmfcalc} and the mass-concentration relation of \protect\cite{child2018halo} }
\label{image:stacked_profile}
\end{figure}

The average mass of the clusters in a stack is,

\begin{equation}
\label{mean_mass_bin}
    \left< \ln M | \ln \lambda \right >_{\rm{bin}} 
    = \int d\ln M P(\ln M|\ln \lambda) \ln M
\end{equation}

which corresponds exactly to what is measured in the case of individual mass estimates of galaxy clusters. We can now compare the excess surface density profile of the average mass in Equation \ref{mean_mass_bin} to the average density profile in Equation \ref{mean_profile}. We show in Figure \ref{image:stacked_profile} these results.

This effect is driven by the mass dependence of the mass concentration relation, because different mass halos contribute at different radial ranges. Such effects whilst small will need to be taken into account in future stacked analyses. Note that beyond 6 Mpc, the lensing signal will become dominated by the 2-halo term, such that these effects will no longer be of importance. Therefore, the effect is important in the intermediate zone of 0.5 Mpc - 5 Mpc. Typical errors on current stacked excess surface mass density measurements are around 20\% compared to the 5\% at 5 Mpc shown in Figure \ref{image:stacked_profile}.

Therefore the results of stacked analysis and that of measurements of individual cluster masses are not directly comparable, however the differences between the two will be small for the reconstructed mass-richness relation and do not account for the observed difference between our results and that of DES. It would be interesting to consider however other correlations between richness, lensing signal, selection effects and contamination of the cluster sample from erroneous cluster detections enter at this level to perturb the relation between the two quantities.

\section{Conclusions}

We have presented a method of estimating galaxy cluster masses from their lensing signals. By restricting our analysis to larger radii, we avoid many systematic effects that impact cluster mass estimation. 

Our calibrated mass-richness relation is in disagreement with the results of the Dark Energy Survey (DES) calibration \citep{mcclintock2019dark} when we use all clusters with richness $\lambda>20$.  This tension disappears when restricting our analysis to higher richness clusters with $\lambda>40$. Estimating masses on individual clusters allow us to constrain additionally the intrinsic scatter between richness and lensing mass. In a full cosmological analysis, this would also allow us to constrain the correlation between richness and lensing mass, which in this study is hidden by the Eddington bias.

The DES year 1 results cosmological constraints from cluster abundances found a significantly lower value of $\Omega_m$ than all other cosmological probes \citep{abbott2020dark}. This analysis uses the mass-richness relation of \cite{mcclintock2019dark}, they find that such a tension would be significantly reduced for a lower richness scaling, $F_l$. Such as that found $F_l=0.981 \pm 0.077$ with the SPT Sunyaev-Zeldovich mass calibration of \cite{bleem2020sptpol}. Therefore the mass-richness relation measured in this work would also reduce the DES year 1 cluster abundance tension with other cosmological probes.

\section*{Acknowledgements}
A portion of the research described in this paper was carried out at the Jet Propulsion Laboratory, California Institute of Technology, under a contract with the National Aeronautics and Space Administration. 

\section*{Data Availability}
The redMaPPer galaxy cluster catalog used in this work may be found at http://risa.stanford.edu/redmapper/. The galaxy shape measurements and photometric redshifts used in the weak lensing analysis are from the first year data of the Hyper Suprime-Cam (HSC) Subaru Strategic Program and may be accessed at this address https://hsc-release.mtk.nao.ac.jp/doc/index.php.



\bibliographystyle{mnras}
\bibliography{example} 



\appendix


\bsp	
\label{lastpage}
\end{document}